# A New Parameter Estimation Algorithm Based on Sub-band Dual Frequency Conjugate LVT

Jing Tian, Wei Cui and Si-liang Wu

*Abstract*— A new parameter estimation algorithm, known as Sub-band Dual Frequency Conjugate LVT (SDFC-LVT), is proposed for the ground moving targets. This algorithm first constructs two sub-band signals with different central frequencies. After that, the two signals are shifted by different values in frequency domain and a new signal is constructed by multiplying one with the conjugate of the other. Finally, Keystone transform and LVT operation are performed on the constructed signal to attain the estimates. The cross-term and the performance of the proposed method are analyzed in detail. Since the equivalent carrier frequency is reduced greatly, the proposed method is capable of obtaining the accurate parameter estimates and resolving the problem of ambiguity which invalidates Keystone transform. It is search-free and can compensate the range walk of multiple targets simultaneously, thereby reducing the computational burden. The effectiveness of the proposed method is demonstrated by both simulated and real data.

I. Introduction

In the complicated modern warfare, the reliable detection of multiple ground moving targets, the acquisition of the estimates of motion parameters, and the re-positioning and imaging of targets are crucial to the achievement of information dominance and have received increasing attention in the radar imaging community [1-5]. Efficient and precise parameter estimation for the signal is a key point for Synthetic aperture radar/ground moving-target imaging (SAR/GMTIm); therefore, it is critical to attain the precise parameter estimates.

Significant research efforts have been devoted to study the multi-targets' parameter estimation and imaging [6-18]. The maximum-likelihood estimation (MLE) [6] was used for parameter estimation and motion compensation. Although MLE is statistically optimal and can obtain high precision, it requires a multi-dimensional joint maximization and is computationally demanding. Time-frequency transforms [7-12] were applied to simultaneous motion compensation and imaging. However, these methods are prone to low-resolution problems as well as cross terms. Then some integration-based methods, including Radon–Wigner transform (RWT) [13, 14], Wigner–Hough transform (WHT) [15], Radon–Ambiguity transform (RAT) [16] and a generalization of the traditional Fourier transform, that is, fractional Fourier transform (FRFT) [17, 18] have been proposed for the parameter estimation of the linear frequency modulated (LFM) signals. However, RWT WHT and RAT are the linear integral detection methods based on the image, which first project the signal onto the time-frequency plane and then search along lines. The FRFT of the signal can be regarded as the projection of its time-frequency distribution onto a rotated frequency axis with a proper rotated angle. It is linear and cross-term free; however, the interference terms around the peaks are very strong because of the order searching, which would affect the precision of parameter estimates. In addition, these existing transformations mentioned above have one important common attribute, that is, exploiting parameter searching operation to obtain parameter estimates.

A novel parameter estimation method for the LFM signals, known as Lv's transform (LVT), has been proposed in [19, 20]. This method breaks through the tradeoff between resolution and cross-terms.

It can obtain accurate parameter estimates without using any searching operation, however, the estimated range of the frequency and chirp rate are $\left[-1/(4T), 1/(4T)\right]$ and $\left[-1/(2T), 1/(2T)\right]$, respectively, where $T$ is the pulse repetition time. The parameters of slow moving targets are located in the corresponding interval, and the accurate parameter estimates can be obtained by LVT. However, the parameters of fast moving targets may be out of the range of the LVT estimator, which means that the estimates of LVT are the mapping of the real value in the corresponding principal value interval.

This paper is aimed at proposing an efficient and accurate parameter estimation method for fast moving target to resolve the aforementioned problem. This method first transforms the received echo into the range frequency domain and then constructs two sub-band signals with different central frequencies. After that, the two signals are shifted by different values and a new synthetic signal is constructed by multiplying one with the conjugate of the other. Finally, Keystone transform and LVT processing are performed on the synthetic signal to attain the estimates. Since the equivalent carrier frequency is reduced greatly, the proposed method is capable of obtaining the accurate parameter estimates and resolving the ambiguity problem which invalidates Keystone transform, especially in the special case when the ambiguity number splits into two numbers.

The remainder of this paper is organized as follows. The signal modeling is given in Section II. The proposed motion parameter estimation approach, namely SDFC-LVT, together with the detailed performance analysis and flowchart of the proposed scheme, are presented in Section III. The performance investigation is given in Section IV. We draw conclusion in Section V, with Appendix appearing in the final Section.

## II. Signal modeling

Suppose that there are $K$ targets in the scene. For simplicity, this section derives the signal model of target $k$. The relevant formulas are valid for other targets due to linearity. The radar usually adopts LFM waveforms, which is expressed as

$$s_T(t,\tau) = a(\tau/T_p)\exp(j\pi\gamma\tau^2)\exp[j2\pi f_c(t+\tau)] \tag{1}$$

where $\tau$ is the fast time, that is, the range time, $t = nT (n = 0,1,\cdots N-1)$ is the slow time, $N$ is the number of coherent integrated pulses, $a(x)$ is the window function and equal to one (for $|x| \leq 1/2$) or zero (otherwise), $T_p$ is the pulse width, $f_c$ is the carrier frequency and $\gamma$ is the chirp rate. Neglecting the high-order components and the change of range within the pulse repetition time, the instantaneous range between radar and the $k$ th target satisfies

$$R_k(t) = R_{0k} - v_{0k}t - \frac{1}{2}a_{0k}t^2 \tag{2}$$

where $R_{0k}$ is the initial range from radar platform to the $k$ th target, $v_{0k}$ and $a_{0k}$ are the radial velocity and the radial acceleration between the target and radar. For SAR-GMTIm system, the instantaneous range can be expressed as [21]

$$R_k(t) = \sqrt{(Vt - v_{ak}t)^2 + (R_{0k} - v_{ck}t)^2} \approx R_{0k} - v_{ck}t + (Vt - v_{ak}t)^2/(2R_{0k}) \tag{3}$$

where $v_{ak}$ and $v_{ck}$ denote along-track and cross-track velocities with projection on the plane of imaging, respectively, and $V$ denotes the velocity of platform.

The received signal after down conversion can be expressed as [22]

$$s_{Rk}(t,\tau) = \sigma_{0k} a\left(\frac{\tau - 2R_k(t)/c}{T_p}\right)\exp\left[-j\frac{4\pi R_k(t)}{\lambda}\right]\exp\left[j\pi\gamma\left(\tau - \frac{2R_k(t)}{c}\right)^2\right] \tag{4}$$

where $\sigma_{0k}$ is the complex reflectivity of the $k$ th target, $\lambda = c/f_c$ is the wavelength and c is the speed of light.

Supposing that $|v_0|/c \ll 1$, then the echo after range compression in the range frequency–slow time $(f,t)$ domain can be expressed as

$$S_{Rk}(t,f) = \sigma'_{0k} a\left(\frac{f}{B}\right) \exp\left[-j\frac{4\pi(f+f_c)}{c} R_k(t)\right] \tag{5}$$

where $B$ is the signal bandwidth.

Substituting (2) into (5) yields

$$S_{Rk}(t,f) = \sigma'_{0k} a\left(\frac{f}{B}\right) \exp\left(-j\frac{4\pi\xi}{\lambda} R_{0k}\right) \exp\left(j\frac{4\pi\xi}{\lambda} v_{0k} t\right) \exp\left(j\frac{2\pi\xi}{\lambda} a_{0k} t^2\right) \tag{6}$$

where $\xi = 1 + f/f_c$.

It should be noted that if we apply LVT to the azimuth signal directly, inaccurate or erroneous estimation of LVT processor may be introduced especially in low signal-to-noise ratio (SNR) scenario, due to the spread energy of the moving target along the range and the azimuth direction. Therefore, the range migration should be corrected first using Keystone transform [23-26]. Besides, the parameters of targets might be out of the range of LVT estimator; therefore, we should also resolve this problem.

III. The proposed method

In what follows, a novel method based on sub-band dual-frequency conjugate processing will be introduced to estimate the parameters for multiple fast moving targets precisely. The detailed process can be achieved as follows.

*A. Derivation*

First, the range spectrum is divided into two parts and two sub-band signals are constructed, as shown in Fig.1. And then the two signals are shifted by $\Delta f_r/2$ and $-\Delta f_r/2$, respectively, where $\Delta f_r = |f_{c1} - f_{c2}|$ denotes the range spectrum separation between the two sub-band signals, in order to make the range dimensional data symmetric in the range frequency domain. The two sub-band signals after shifting can be shown as

$$s_{Rk\_part1}(t,\tau) = \sigma_{0k}G_1 \text{sinc}\left[\pi\frac{B}{2}\left(\tau - \frac{2R_k(t)}{c}\right)\right]\exp\left[-j\frac{4\pi}{c}\left(f_c - \frac{\Delta f_r}{2}\right)R_k(t)\right] \quad (7)$$

$$s_{Rk\_part2}(t,\tau) = \sigma_{0k}G_2 \text{sinc}\left[\pi\frac{B}{2}\left(\tau - \frac{2R_k(t)}{c}\right)\right]\exp\left[-j\frac{4\pi}{c}\left(f_c + \frac{\Delta f_r}{2}\right)R_k(t)\right] \quad (8)$$

where $G_1$ and $G_2$ are the range compression gain of the two sub-signals and satisfy $G_1 = G_2 = \sqrt{B\tau}/2$.

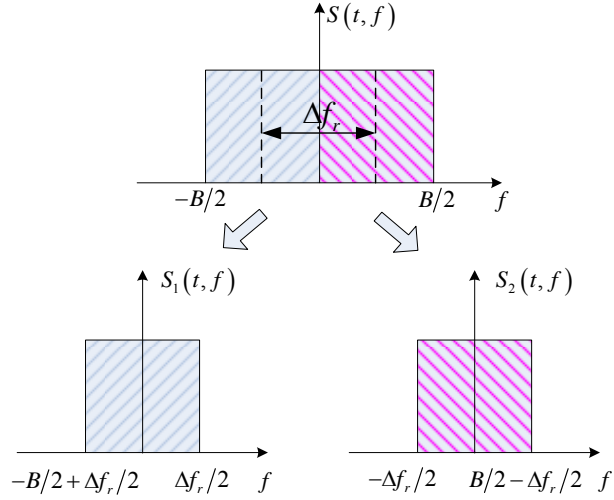

Fig.1. sub-band signal construction in range frequency domain.

Multiplying $s_{Rk\_part2}(t,\tau)$ with the conjugate of $s_{Rk\_part1}(t,\tau)$, the auto-term of the new signal can be expressed as

$$\begin{aligned} s_{comp}(t,\tau) &= s_{Rk\_part2}(t,\tau) \cdot \text{conj}\left(s_{Rk\_part1}(t,\tau)\right) \\ &= \sigma_{0k}^2 G_1 G_2 \text{sinc}^2\left[\pi\frac{B}{2}\left(\tau - \frac{2R_k(t)}{c}\right)\right]\exp\left[-j\frac{4\pi}{c}\Delta f_r R_k(t)\right] \end{aligned} \quad (9)$$

It can be seen from (9) that, the equivalent carrier frequency of the signal $s_{comp}(t,\tau)$ turns into $\Delta f_r$, which satisfies $\Delta f_r \ll f_c$. The maximum unambiguous velocity is $|v_{max}| = c/(4f_c T)$ without sub-band dual-frequency conjugate processing, while it becomes $|v'_{max}| = c/(4\Delta f_r T)$ after using the proposed method. Although the velocity of target is ambiguous for the traditional methods, the range walk of the synthetic signal can be corrected through Keystone transform as long as the velocity and acceleration satisfy $|v_{0k}+a_{0k}NT|\leq|v'_{\max}|$, which means

$$|v_{0k}+a_{0k}NT|\leq \frac{c}{4\Delta f_r T} \tag{10}$$

After the aforementioned pre-processing, Keystone transform [23-26] and LVT [19, 20] operation are performed on the constructed signal (shown in (9)) in the range dimension and azimuth dimension, respectively, to attain the estimates. And the input parameter pair of the LVT estimator turns into $(2v_{0k}\Delta f_r/c,\ 2a_{0k}\Delta f_r/c)$. To ensure the accurate parameter estimates can be obtained, the following inequality should hold:

$$\begin{cases}\left|\dfrac{2v_{0k}\Delta f_r}{c}\right|\leq \dfrac{1}{4T}\\ \left|\dfrac{2a_{0k}\Delta f_r}{c}\right|\leq \dfrac{1}{2T}\end{cases} \tag{11}$$

The range of velocity and acceleration obtained by the LVT estimator without and with using the proposed method are shown in Fig.2. The left oblique line denotes the output range of LVT estimator without using the proposed method. After applying the proposed method, the range of parameters expands to the region indicated by the right oblique line.

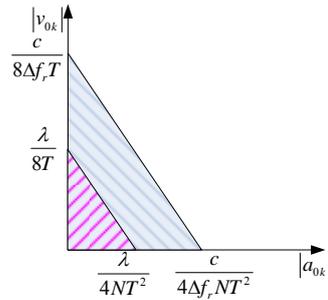

Fig.2. the range of velocity and acceleration without and with using the proposed method.

B. *Analysis of the Cross-term*

Following the aforementioned processing, each target is completely accumulated and corresponding to a sole peak in the velocity-acceleration distribution plane. This will benefit the estimation of the velocity and the acceleration. However, the cross-term appears between different targets and may influence the precision of parameter estimates. Thus, we analyze characteristics of the cross-term below. According to the definition of the synthetic signal, the cross-term can be denoted as

$$s_{comp\_cross}(t,\tau) = \sum_{i=1}^{K-1}\sum_{j=i+1}^{K}\left[s_{Ri\_part2}(t,\tau)\cdot\text{conj}\left(s_{Rj\_part1}(t,\tau)\right)+s_{Rj\_part2}(t,\tau)\cdot\text{conj}\left(s_{Ri\_part1}(t,\tau)\right)\right]$$

$$= \sum_{i=1}^{K-1}\sum_{j=i+1}^{K}\left\{\sigma_{0i}\sigma_{0j}G_iG_j\text{sinc}\left[\pi\frac{B}{2}\left(\tau-\frac{2R_i(t)}{c}\right)\right]\text{sinc}\left[\pi\frac{B}{2}\left(\tau-\frac{2R_j(t)}{c}\right)\right]\right.$$

$$\cdot\exp\left[j\frac{4\pi}{c}f_{c1}R_i(t)\right]\exp\left[-j\frac{4\pi}{c}f_{c2}R_j(t)\right]$$

$$+\sigma_{0i}\sigma_{0j}G_iG_j\text{sinc}\left[\pi\frac{B}{2}\left(\tau-\frac{2R_i(t)}{c}\right)\right]\text{sinc}\left[\pi\frac{B}{2}\left(\tau-\frac{2R_j(t)}{c}\right)\right]$$

$$\left.\cdot\exp\left[j\frac{4\pi}{c}f_{c1}R_j(t)\right]\exp\left[-j\frac{4\pi}{c}f_{c2}R_i(t)\right]\right\}$$

$$= \sum_{i=1}^{K-1}\sum_{j=i+1}^{K}\left\{\begin{array}{l}\sigma_{0i}\sigma_{0j}G_iG_j\text{sinc}\left[\pi\frac{B}{2}\left(\tau-\frac{2R_i(t)}{c}\right)\right]\text{sinc}\left[\pi\frac{B}{2}\left(\tau-\frac{2R_j(t)}{c}\right)\right]\\ \cdot\exp\left[j\frac{4\pi}{c}(f_{c1}-f_{c2})R_j(t)\right]\\ \cdot\left\{\exp\left[j\frac{4\pi}{c}f_{c1}(R_i(t)-R_j(t))\right]+\exp\left[-j\frac{4\pi}{c}f_{c2}(R_i(t)-R_j(t))\right]\right\}\end{array}\right\}$$

(12)

The signal $s_{comp\_cross}(t,\tau)$ in the range frequency and azimuth time domain is formulated as

$$s_{comp\_cross}(t,f) = \sum_{i=1}^{K-1}\sum_{j=i+1}^{K}\sigma_{0i}\sigma_{0j}G_iG_j\frac{\sin\left[\frac{2\pi}{c}(R_j(t)-R_i(t))\left(\frac{B}{2}-|f|\right)\right]}{\sin\left[\frac{2\pi}{c}(R_j(t)-R_i(t))\frac{f_s}{N}\right]}$$

$$\cdot\exp\left[j\frac{4\pi}{c}(f_{c1}-f_{c2})R_j(t)\right]\exp\left[-j\frac{2\pi}{c}f(R_i(t)+R_j(t))\right]$$

$$\cdot\left\{\exp\left[j\frac{4\pi}{c}f_{c1}(R_i(t)-R_j(t))\right]+\exp\left[-j\frac{4\pi}{c}f_{c2}(R_i(t)-R_j(t))\right]\right\}$$

(13)

Performing Keystone transform on the signal $s_{comp\_cross}(t,f)$, that is, substituting the scaling factor $t = \Delta f_r t_a/(\Delta f_r + f)$ into $s_{comp\_cross}(t,f)$, and adopting the inverse fast Fourier transform (IFFT) on $s_{comp\_cross}(t_a,f)$ with respect to $f$, the results are listed below and the detailed analysis can be referred to Appendix A.

(1) For $R_j(t) = R_i(t)$,

$$s_{comp\_cross}(t_a,\tau) = 2\sigma_{0i}\sigma_{0j}G_iG_j\operatorname{sinc}^2\left[\pi\frac{B}{2}\left(\tau-2\frac{R_{0i}+\frac{1}{2}a_{0i}t^2}{c}\right)\right]\exp\left[-j\frac{4\pi}{c}\Delta f_r R_i(t_a)\right] \quad (14)$$

(2) For $R_j(t) \neq R_i(t)$,

$$s_{comp\_cross}(t_a,\tau) = \sum_{i=1}^{K-1}\sum_{j=i+1}^{K}\sigma_{0i}\sigma_{0j}G_iG_j\left[\Delta R_1(t_a,\tau)+\Delta R_2(t_a,\tau)\right] \quad (15)$$

$$\Delta R_1(t_a,\tau) = \operatorname{sinc}^2\left[\pi\frac{B}{2}\left(\tau-\frac{R_{0i}+R_{0j}+\frac{1}{2}(a_{0i}+a_{0j})t_a^2}{c}-\frac{(f_{c1}+f_{c2})\left[(v_{0i}-v_{0j})t_a+(a_{0i}-a_{0j})t_a^2\right]}{c\Delta f_r}\right)\right]$$
$$\cdot\exp\left[-j\frac{2\pi}{c}(f_{c1}+f_{c2})\left[R_i(t_a)-R_j(t_a)\right]\right]\exp\left[-j\frac{2\pi}{c}\Delta f_r\left[R_i(t_a)+R_j(t_a)\right]\right]$$

(15A)

$$\Delta R_2(t_a,\tau) = \operatorname{sinc}^2\left[\pi\frac{B}{2}\left(\tau-\frac{R_{0i}+R_{0j}+\frac{1}{2}(a_{0i}+a_{0j})t_a^2}{c}+\frac{(f_{c1}+f_{c2})\left[(v_{0i}-v_{0j})t_a+(a_{0i}-a_{0j})t_a^2\right]}{c\Delta f_r}\right)\right]$$
$$\cdot\exp\left[j\frac{2\pi}{c}(f_{c1}+f_{c2})\left[R_i(t_a)-R_j(t_a)\right]\right]\exp\left[-j\frac{2\pi}{c}\Delta f_r\left[R_i(t_a)+R_j(t_a)\right]\right]$$

(15B)

It is obvious that when the parameter values of two targets are equal completely, i.e., $R_j(t) = R_i(t)$ holds, the energy of the cross-term is strong in (14). And in this situation, the cross-term turns out to be the auto-term and the parameter estimates can be obtained. However, when $R_j(t) \neq R_i(t)$, the coupling between $t_a$ and $f$ cannot be eliminated, thereby defocusing the energy of cross-term and reducing the amplitude of output signal greatly. The detailed analysis can be referred to Appendix A. According to the analysis above, the auto-terms corresponding to the targets turn out to be the distinct points, whereas the cross-terms generated by different targets are still distributed on the LVT plane. Their energy can be ignored compared to the peaks of auto-terms. That is to say, the cross-term is relatively suppressed since LVT greatly strengthens the energy of the auto-term instead of suppressing the cross-term directly.

*C. Analysis of SNR*

It is necessary to analyze the output SNR after the sub-band dual-frequency conjugate processing to characterize the performance of the pre-processor. In the absence of noise, i.e., $x(t,\tau) = s_{comp\_cross}(t,\tau)$, the test statistic of the proposed processor at the maximum point $\tau_0 = \frac{2R_k(t)}{c}$ is denoted by $\text{SDFC}_s(\tau_0)$. In the presence of noise, $x(t,\tau) = s_{comp\_cross}(t,\tau) + n(t,\tau)$, the test statistic at $\tau_0$ is a random variable and is denoted as $\text{SDFC}_{s+n}(\tau_0)$. As a result, the output SNR is defined as [27, 28]

$$\text{SNR}_{\text{SDFC}} = \frac{|\text{SDFC}_s(\tau_0)|^2}{\text{var}\{\text{SDFC}_{s+n}(\tau_0)\}} \tag{16}$$

where $\text{var}\{\cdot\}$ denotes the variance of its argument. The input SNR and the SNR after range compression are denoted as $\text{SNR}_{\text{in}} = \sigma_{0k}^2/V^2$ and $\text{SNR}_{\text{PC}} = B\tau \cdot \text{SNR}_{\text{in}}$, respectively, where $V^2$ is the variance of an additive zero-mean Gaussian noise. The output SNR is shown below and the detailed analysis is derived in Appendix B:

$$\text{SNR}_{\text{SDFC}} = \frac{\left(\frac{\sigma_{0k}^2}{4}B\tau\right)^2}{\frac{V^4}{4} + \frac{\sigma_{0k}^2}{4}B\tau V^2} = \frac{\text{SNR}_{\text{PC}}^2}{4 + 4\text{SNR}_{\text{PC}}} \tag{17}$$

This relationship reveals the presence of a threshold effect. In fact, at high SNR after range compression ($\text{SNR}_{\text{PC}} \gg 1$), the previous expression can be approximated by $\text{SNR}_{\text{SDFC}} \approx \text{SNR}_{\text{PC}}/4$ and then there is a 6-dB loss due to the nonlinearity, compared with the original range compression process. Conversely, at low SNR ($\text{SNR}_{\text{PC}} \ll 1$), the SNR after sub-band dual-frequency conjugate processing could be even worse than $\text{SNR}_{\text{PC}}$. We can define the threshold of the pre-processor as the interception point between the two limiting behaviors corresponding to the two cases of high and low SNR, obtaining a threshold value for $\text{SNR}_{\text{PC}}$ equal to 1.

Based on (17) and the result in [19], the SNR after LVT operation is bounded by

$$\text{SNR}_{\text{LVT}} > \frac{N^2(B\tau)^4 \text{SNR}_{\text{in}}^4}{8N(B\tau)^3 \text{SNR}_{\text{in}}^3 + (8N+32)(B\tau)^2 \text{SNR}_{\text{in}}^2 + 64B\tau\text{SNR}_{\text{in}} + 32} \tag{18}$$

*D. Analysis of Accuracy*

Although the proposed SDFC-LVT is an accurate parameter estimation method without any searching, it also inevitably suffers from representation errors in the presence of noise. The performance of parameter representation is usually evaluated by a perturbation analysis method [19, 22]. In the presence of signal plus noise, the received signal can be written as the sum of a useful term $s_{comp\_cross}(t,\tau)$ depending only on the signal, plus a perturbation $\delta x(t,\tau)$ depending on the noise and on its interaction with the signal. Denote by $\delta v_0$ and $\delta a_0$ the errors of parameters $v_0$ and $a_0$. Then the signal after the sub-band dual-frequency conjugate processing is decomposed as

$$x(t,\tau) = s_{comp\_cross}(t,\tau) + \delta x(t,\tau) \tag{19}$$

After LVT operation in the azimuth direction, we can obtain

$$\text{LVT}_x(v,a) = \text{LVT}_s(v,a) + \delta\text{LVT}_x(v,a) \tag{20}$$

The result of the perturbation analysis in [19] is employed and the variances of the estimates of velocity and acceleration are bounded by

$$\sigma_v^2 < \frac{c^2\left(147N^3 + 36q^2N^2\right)\left(1 + B\tau\text{SNR}_{in}\right)}{\pi^2 \Delta f_r^2 \left(98N^4 + 72q^4\right)(B\tau)^2 \text{SNR}_{in}^2} \tag{21}$$

$$\sigma_a^2 < \frac{588c^2h^2\left(1 + B\tau\text{SNR}_{in}\right)}{2\pi^2 \Delta f_r^2 N (B\tau)^2 \text{SNR}_{in}^2} \tag{22}$$

where $q$ is the number of sampling points of a constant time-delay.

The detailed flowchart of the proposed method is illustrated in Fig.3.

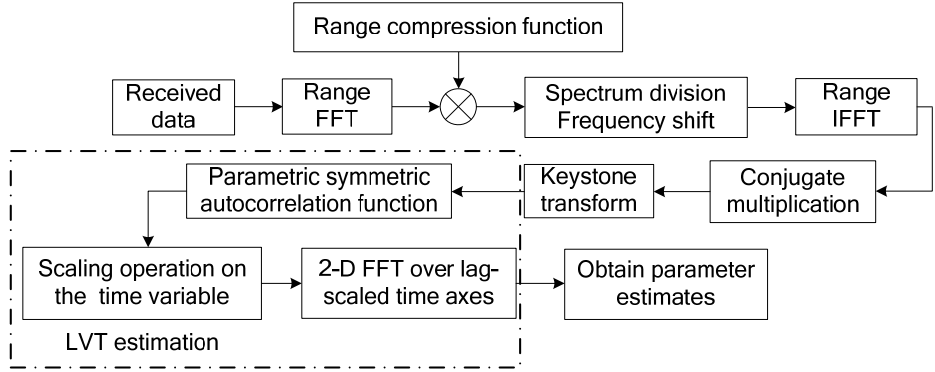

Fig.3. the flowchart of the proposed method.

IV. Experimental results

In this section, some results with simulated and real data [29] are presented to validate the proposed parameter estimation algorithm.

*A. Simulated Data*

First, echoes of two targets with high speed and low speed respectively are generated to prove the validity of the proposed method. The parameters of targets and the simulated parameters are listed in TABLE 1 and TABLE 2.

TABLE 1 INITIAL POSITIONS AND MOTION PARAMETERS OF THE TWO TARGETS

|  | Initial range (km) | Radial velocity (m/s) | Radial acceleration (m/s$^2$) |
| --- | --- | --- | --- |
| Target 1 | 15.3 | 197.87 | 4.88 |
| Target 2 | 15.3 | 70.92 | 4.88 |

TABLE 2 SYSTEM PARAMETERS OF RADAR

| System parameters (Unit) | Values |
| --- | --- |
| Carrier frequency (GHz) | 1 |
| Pulse width (us) | 4 |
| Bandwidth (MHz) | 15 |
| Sampling frequency (MHz) | 37.5 |
| Pulse repetition interval (us) | 500 |
| Coherent integrated pulses | 2048 |

Fig.4 shows the result of the proposed SDFC-LVT. Fig. 4(a) shows the trajectory of the new synthetic signal after range compression. It is obvious that the signal energy of two targets spreads over several range cells. We perform Keystone transform to correct the range cell migration (RCM) and obtain the result in Fig. 4(b). It is observed that the RCMs of two targets are eliminated effectively. After LVT, as shown in Fig. 4(c), the targets are well focused. The parameters of targets with the values of $R_{01}=15.3$ km, $v_{01}=197.8711$ m/s, $a_{01}=4.8828$ m/s$^2$, $R_{02}=15.3$ km, $v_{02}=70.9180$ m/s and $a_{02}=4.8828$ m/s$^2$, respectively, are also obtained. In addition, the cross-terms generated by the two targets are still distributed on the LVT plane. Since Keystone transform cannot remove the coupling between the time variable and the lag variable, the energy of the cross-term after LVT processing cannot be concentrated completely.

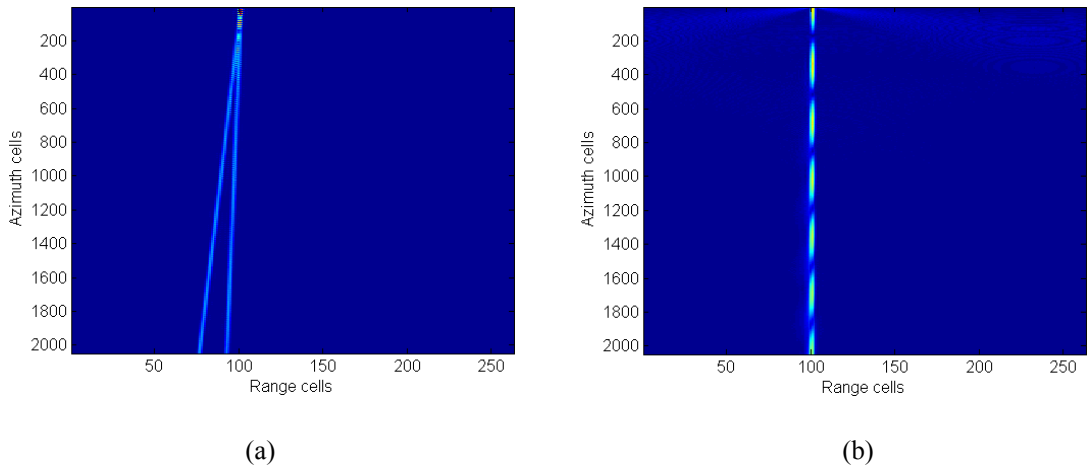

(a)          (b)

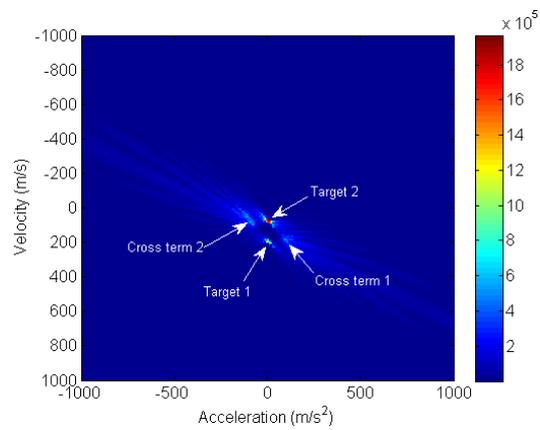

(c)

Fig.4. Simulation results of the proposed method. (a) Trajectory of the synthetic signal after range compression. (b) Trajectory after Keystone transform. (c) Result of LVT.

In the following, comparison of the performance between the proposed SDFC-LVT and the second-order radon Fourier transform (sRFT) [30, 31] is performed. The signal is embedded in complex white Gaussian noise and Monte Carlo simulation results quantify the root-mean-square (RMS) errors of the estimates of the targets with respect to the SNR, in which 500 trials are used for each SNR. The amplitudes of the two targets satisfy $\sigma_{01}^2 = 2\sigma_{02}^2$. Figs. 5(a) and 5(b) show the RMS errors of the acceleration and velocity estimates for target 1, respectively. And Figs. 6(a) and 6(b) show the RMS errors of the acceleration and velocity estimates for target 2, respectively.

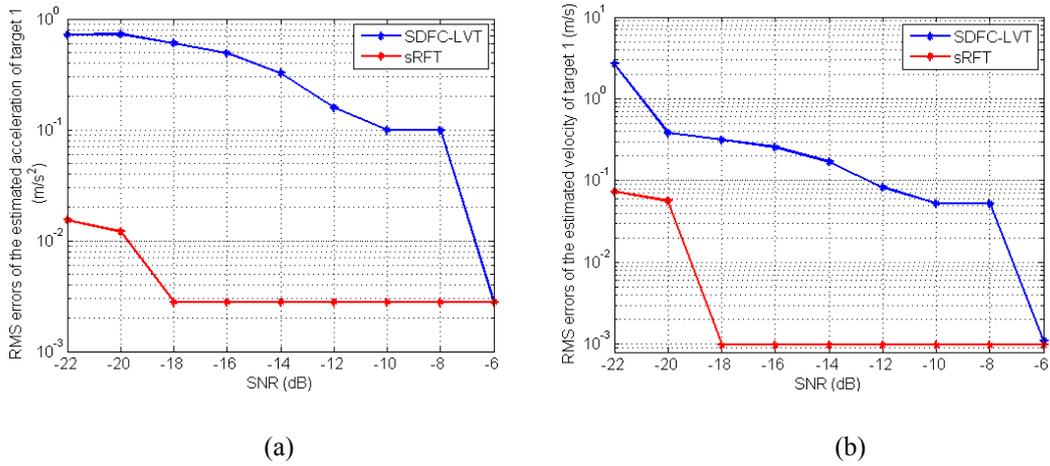

(a)          (b)

Fig.5. RMS errors of (a) acceleration and (b) velocity against input SNRs via the SDFC-LVT method and the sRFT method for target 1.#

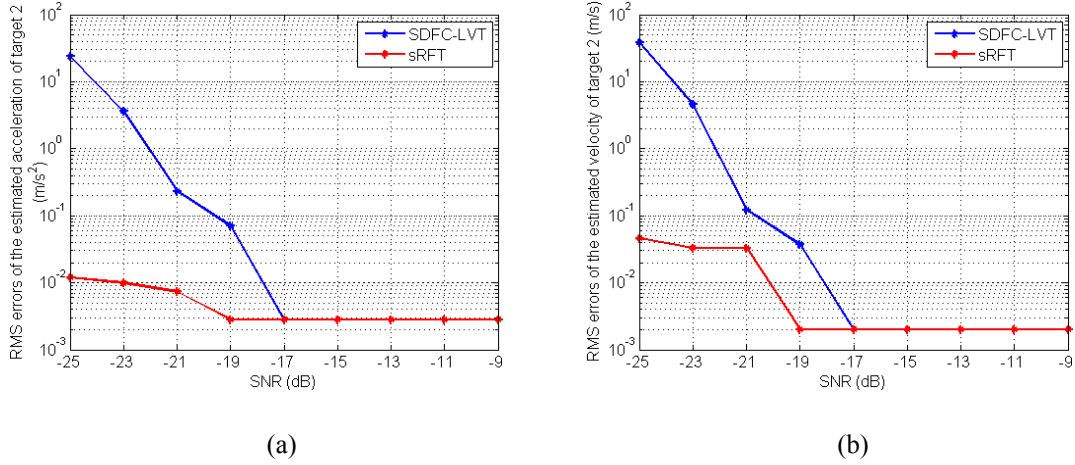

(a)                  (b)

Fig.6. RMS errors of (a) acceleration and (b) velocity against input SNRs via the SDFC-LVT method and the sRFT method for target 2.

As shown in Fig.5 and Fig.6, the performance of the proposed SDFC-LVT method is not as good as that of the sRFT method because the cross-term results in the degradation of the performance with the decreasing of SNR for the proposed method, however, its performance is still acceptable in realistic applications [32]. In addition, the proposed SDFC-LVT method can simultaneously estimate the parameters of slow and fast moving targets precisely. Since the sRFT method needs three-dimensional searching while the SDFC-LVT method is search-free and can be easily implemented by using the complex multiplications and FFT based on the scaling principle, the SDFC-LVT method has much lower computation complex. Based on above analysis, we know that, compared to the sRFT method, the proposed SDFC-LVT method obtains a good balance between the computational cost and the estimation performance, making it more suitable for realistic applications.

*B. Real Data*

Part of the RADARSAT-1 Vancouver scene data were selected to verify our proposed method and analysis. The system parameters of these data are given in TABLE 3 and the proposed procedure is performed on the selected target (labeled in the Fig. 7(a)). Fig. 7(b) shows the result of new synthetic signal after range compression, from which it can be seen that the large RCM exists in the plane. Fig. 7(c) shows the result after Keystone transform, from which it can be seen that the large RCM is eliminated completely. After LVT, as shown in Fig. 7(d), the target is well focused. And the velocity and acceleration with the value of $-207.7265\,\text{m/s}$ and $50.4445\,\text{m/s}^2$, respectively, are also estimated. The corresponding frequency and chirp rate are equal to $-7344.8120\,\text{Hz}$ and $1783.6215\,\text{Hz/s}$, respectively, which are consistent with the results obtained by the parameter estimation method in [33], thereby verifying the effectiveness of the new approach. The result of correcting the trajectory of received signal after range compression with the estimated parameters is shown in Fig. 7(e), from which it can be seen that the large RCM is removed completely. The result of applying Keystone transform on the received signal after range compression directly is shown in Fig. 7(f), from which it can be seen that the large RCM cannot be eliminated completely because of the velocity ambiguity.

It can also be concluded that the proposed SDFC-LVT method can avoid the influence of Doppler ambiguity on Keystone transform and LVT estimator.

TABLE 3 SYSTEM PARAMETERS FOR RADARSAT DATA

| System parameters | Values |
| --- | --- |
| Carrier frequency (GHz) | 5.3 |
| Range bandwidth (MHz) | 30.116 |
| Pulse repetition frequency (Hz) | 1256.98 |
| Range sampling frequency (MHz) | 32.317 |
| Pulse width (us) | 41.74 |
| Doppler centriod frequency (Hz) | -6900 |
| Azimuth chirp rate (Hz/s) | 1733 |

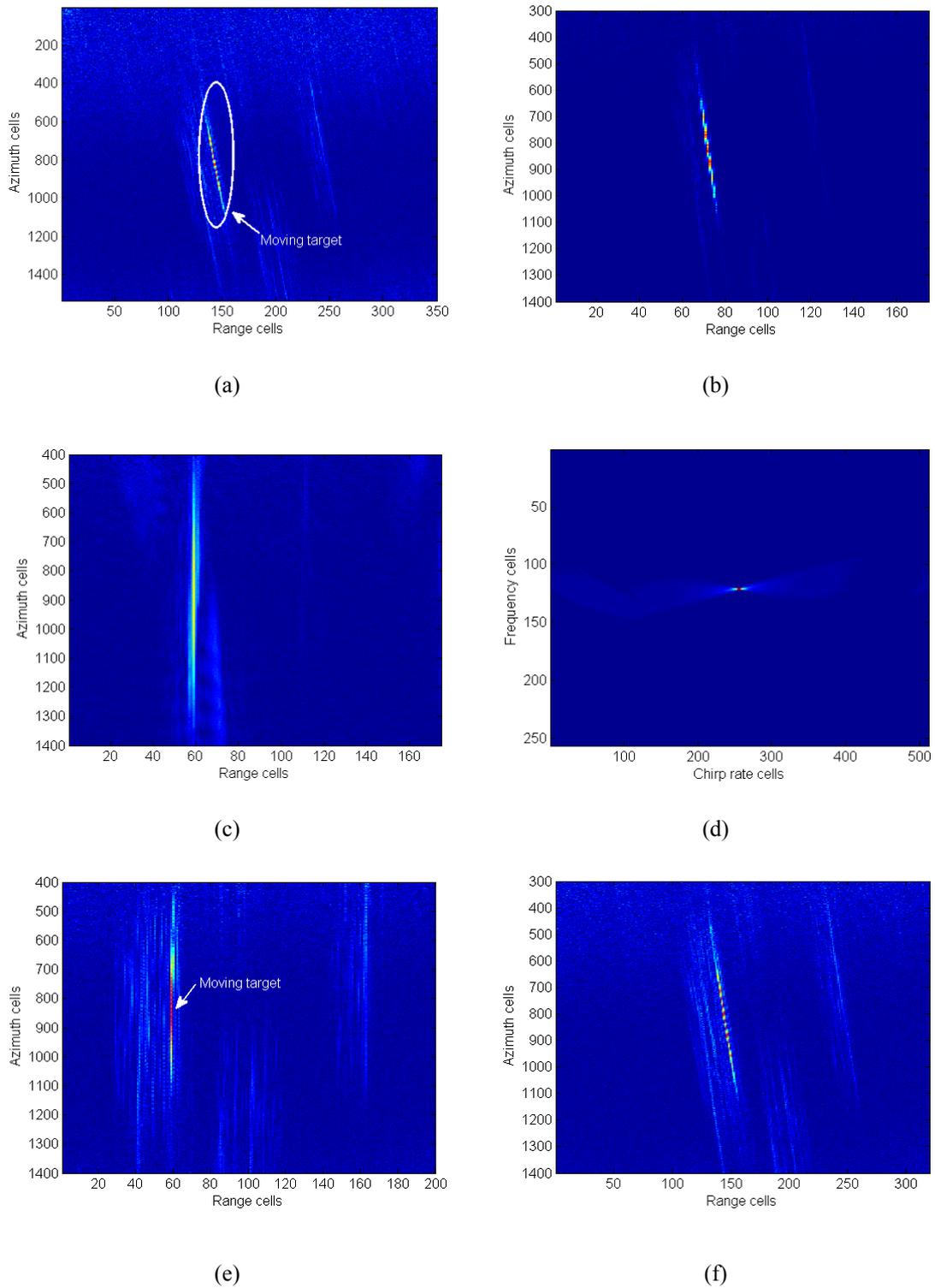

Fig. 7. Result of the Real data via the proposed method. (a) Trajectory after range compression. (b) Trajectory of the new synthetic signal after range compression. (c) Trajectory after Keystone transform. (d) Result of LVT. (e) Trajectory after range compression and motion compensation. (f) Result of applying Keystone transform on the received signal after range compression directly.

V.  Conclusions

This paper has introduced a parametric estimation method, known as SDFC-LVT, for the ground moving targets. It is capable of obtaining the motion parameters with high precision and also computationally efficient. A new synthetic signal is constructed by the essential operations, including spectrum division, frequency shifting, and conjugate multiplications. After that, Keystone transform and LVT operator are performed on this new signal. This method can unambiguously estimate motion parameters. The characteristics of SDFC-LVT include the following: (a) It is a nonsearching method; (b) there is no need to compensate the first-order term separately for each target, resulting in reduced computational burden; and (c) the method can be performed without prior-knowledge of targets, which can decrease the demand for the inertial navigation system. The simulated and real data processing results have validated the effectiveness of the proposed algorithm.

APPENDIX A

In this proposed parameter estimation algorithm, each signal component generates an auto-term and each pair of signal components generates a cross-term. For given $i \in [1, K-1]$ and $j \in [i+1, K]$, the cross-term $s_{comp\_cross}(t, \tau)$ can be expressed as

$$s_{comp\_cross}(t,\tau) = \sum_{i=1}^{K-1} \sum_{j=i+1}^{K} \left[ s_{Ri\_part2}(t,\tau) \cdot \text{conj}(s_{Rj\_part1}(t,\tau)) + s_{Rj\_part2}(t,\tau) \cdot \text{conj}(s_{Ri\_part1}(t,\tau)) \right]$$

$$= \sum_{i=1}^{K-1} \sum_{j=i+1}^{K} \left\{ \begin{array}{l} \sigma_{0i}\sigma_{0j}G_iG_j \text{sinc}\left[\pi\frac{B}{2}\left(\tau-\frac{2R_i(t)}{c}\right)\right]\text{sinc}\left[\pi\frac{B}{2}\left(\tau-\frac{2R_j(t)}{c}\right)\right] \\ \cdot \exp\left[j\frac{4\pi}{c}(f_{c1}-f_{c2})R_j(t)\right] \\ \cdot \left\{\exp\left[j\frac{4\pi}{c}f_{c1}(R_i(t)-R_j(t))\right] + \exp\left[-j\frac{4\pi}{c}f_{c2}(R_i(t)-R_j(t))\right]\right\} \end{array} \right\} \quad (23)$$

Case 1: $R_j(t) = R_i(t)$

The signal $s_{comp\_cross}(t,\tau)$ in the range frequency and azimuth time domain is formulated as

$$s_{comp\_cross}(t,f) = \sum_{i=1}^{K-1}\sum_{j=i+1}^{K} \sigma_{0i}\sigma_{0j}G_iG_j \frac{4}{B}\left[1-\frac{|f|}{B/4}\right]\exp\left[-j2\pi(f+\Delta f_r)\frac{2R_i(t)}{c}\right] \quad (24)$$

Performing Keystone transform on the signal $s_{comp\_cross}(t,f)$, that is, substituting the scaling factor $t = \Delta f_r t_a/(\Delta f_r + f)$ into $s_{comp\_cross}(t,f)$ yields

$$\begin{aligned}s_{comp\_cross}(t_a,f) \approx &\sum_{i=1}^{K-1}\sum_{j=i+1}^{K} \sigma_{0i}\sigma_{0j}G_iG_j \frac{4}{B}\left[1-\frac{|f|}{B/4}\right]\exp\left[-j2\pi\Delta f_r \frac{2R_i(t_a)}{c}\right]\\ &\cdot \exp\left[-j2\pi f \frac{2}{c}\left(R_{0i}+\frac{1}{2}a_{0i}t_a^2\right)\right]\end{aligned} \quad (25)$$

In (25), the expression $\dfrac{1}{\Delta f_r + f} = \dfrac{1}{\Delta f_r}\left[1+\sum_{i=1}^{\infty}\left(-\dfrac{f}{\Delta f_r}\right)^i\right]$ [34] and the approximation $f < \Delta f_r$ is used.

Then adopting IFFT on $s_{comp\_cross}(t_a,f)$ with respect to $f$, we can obtain

$$s_{comp\_cross}(t_a,\tau) = \sum_{i=1}^{K-1}\sum_{j=i+1}^{K} 2\sigma_{0i}\sigma_{0j}G_iG_j \operatorname{sinc}^2\left[\pi\frac{B}{2}\left(\tau - 2\frac{R_{0i}+\frac{1}{2}a_{0i}t^2}{c}\right)\right]\exp\left[-j\frac{4\pi}{c}\Delta f_r R_i(t_a)\right] \quad (26)$$

Case 2: $R_j(t) \neq R_i(t)$

The signal $s_{comp\_cross}(t,\tau)$ in the range frequency and azimuth time domain is further expressed as

$$\begin{aligned}s_{comp\_cross}(t,f) = &\sum_{i=1}^{K-1}\sum_{j=i+1}^{K} \sigma_{0i}\sigma_{0j}G_iG_j \frac{\sin\left[\frac{2\pi}{c}\left(R_j(t)-R_i(t)\right)\left(\frac{B}{2}-|f|\right)\right]}{\sin\left[\frac{2\pi}{c}\left(R_j(t)-R_i(t)\right)\frac{f_s}{N}\right]}\\ &\cdot 2\cos\left[\frac{2\pi}{c}\left(R_i(t)-R_j(t)\right)(f_{c1}+f_{c2})\right]\\ &\cdot \exp\left[j\frac{2\pi}{c}(f_{c1}-f_{c2}-f)(R_i(t)+R_j(t))\right]\end{aligned} \quad (27)$$

Performing Keystone transform on the signal $s_{comp\_cross}(t,f)$, that is, substituting the scaling factor $t = \Delta f_r t_a/(\Delta f_r + f)$ into $s_{comp\_cross}(t,f)$ yields

$$s_{comp\_cross}(t_a, f) \approx \sum_{i=1}^{K-1} \sum_{j=i+1}^{K} \sigma_{0i}\sigma_{0j} G_i G_j P_1 P_2 \exp\left\{-j\frac{2\pi}{c} f\left[R_{0i} + R_{0j} + \frac{1}{2}(a_{0i} + a_{0j})t_a^2\right]\right\}$$
$$\cdot \exp\left\{-j\frac{2\pi}{c}\Delta f_r \left[R_{0i} + R_{0j} - (v_{0i} + v_{0j})t_a - \frac{1}{2}(a_{0i} + a_{0j})t_a^2\right]\right\} \quad (28)$$

$$P_1(t_a, f) = \frac{\sin\left\{\frac{2\pi}{c}\left[R_{0j} - R_{0i} - (v_{0j} - v_{0i})\left(1 - \frac{f}{\Delta f_r}\right)t_a - \frac{1}{2}(a_{0j} - a_{0i})\left(1 - \frac{f}{2\Delta f_r}\right)t_a^2\right]\left(\frac{B}{2} - |f|\right)\right\}}{\sin\left\{\frac{2\pi}{c}\left[R_{0j} - R_{0i} - (v_{0j} - v_{0i})\left(1 - \frac{f}{\Delta f_r}\right)t_a - \frac{1}{2}(a_{0j} - a_{0i})\left(1 - \frac{f}{2\Delta f_r}\right)t_a^2\right]\frac{f_s}{N}\right\}} \quad (28A)$$

$$P_2(t_a, f) = 2\cos\left\{\frac{2\pi}{c}\left[R_{0i} - R_{0j} - (v_{0i} - v_{0j})\left(1 - \frac{f}{\Delta f_r}\right)t_a - \frac{1}{2}(a_{0i} - a_{0j})\left(1 - \frac{f}{2\Delta f_r}\right)t_a^2\right](f_{c1} + f_{c2})\right\} \quad (28B)$$

As seen in (28A), the envelope of $P_1(t_a, f)$ along the approximate straight-line $R_{0j} - R_{0i} - (v_{0j} - v_{0i})t_a - \frac{1}{2}(a_{0j} - a_{0i})t_a^2 \approx 0$ has a triangular shape varying with $f$. Using the Fourier transform of triangular function and performing IFFT on (28A) with respect to $f$ yields

$$Q_1(t_a, \tau) = IFFT_f\left[P_1(t_a, f)\right] \approx \operatorname{sinc}^2\left[\pi\frac{B}{2}\tau\right] \quad (29)$$

Then performing IFFT on (28B) with respect to $f$ yields

$$Q_2(t_a, \tau) = IFFT_f\left[P_2(t_a, f)\right]$$
$$= \left\{\delta\left[\tau - \frac{(f_{c1} + f_{c2})\left[(v_{0i} - v_{0j})t_a + (a_{0i} - a_{0j})t_a^2\right]}{c\Delta f_r}\right] + \delta\left[\tau + \frac{(f_{c1} + f_{c2})\left[(v_{0i} - v_{0j})t_a + (a_{0i} - a_{0j})t_a^2\right]}{c\Delta f_r}\right]\right\} \quad (30)$$
$$\cdot \exp\left\{-j2\pi \frac{R_i(t_a) - R_j(t_a)}{(v_{0i} - v_{0j})t_a + (a_{0i} - a_{0j})t_a^2}\Delta f_r \tau\right\}$$

Combining (29) with (30) and utilizing the convolution property, we can represent the cross-term in the range time and azimuth time domain as

$$s_{comp\_cross}(t_a, \tau) = \sum_{i=1}^{K-1}\sum_{j=i+1}^{K}\sigma_{0i}\sigma_{0j}G_iG_j\left[\Delta R_1(t_a, \tau) + \Delta R_2(t_a, \tau)\right] \quad (31)$$

$$\Delta R_1(t_a, \tau) = \operatorname{sinc}^2\left[\pi\frac{B}{2}\left(\tau - \frac{R_{0i} + R_{0j} + \frac{1}{2}(a_{0i} + a_{0j})t_a^2}{c} - \frac{(f_{c1} + f_{c2})\left[(v_{0i} - v_{0j})t_a + (a_{0i} - a_{0j})t_a^2\right]}{c\Delta f_r}\right)\right]$$
$$\cdot \exp\left[-j\frac{2\pi}{c}(f_{c1} + f_{c2})\left[R_i(t_a) - R_j(t_a)\right]\right]\exp\left[-j\frac{2\pi}{c}\Delta f_r\left[R_i(t_a) + R_j(t_a)\right]\right]$$

$$\Delta R_2(t_a,\tau) = \text{sinc}^2\left[\pi\frac{B}{2}\left(\tau - \frac{R_{0i}+R_{0j}+\frac{1}{2}(a_{0i}+a_{0j})t_a^2}{c} + \frac{(f_{c1}+f_{c2})\left[(v_{0i}-v_{0j})t_a+(a_{0i}-a_{0j})t_a^2\right]}{c\Delta f_r}\right)\right]$$

(31A)

$$\cdot \exp\left[j\frac{2\pi}{c}(f_{c1}+f_{c2})\left[R_i(t_a)-R_j(t_a)\right]\right]\exp\left[-j\frac{2\pi}{c}\Delta f_r\left[R_i(t_a)+R_j(t_a)\right]\right]$$

(31B)

Equations (31), (31A) and (31B) are the derivations of the results in Section III. It can be seen from (31A) and (31B) that, the cross-term cannot be accumulated as the auto-term because the trajectory is not a straight line, with its energy spreading along range.

## APPENDIX B

Let us consider the signal after sub-band dual-frequency conjugate processing

$$x(t,\tau) = \left[s_{Rk\_part2}(t,\tau)+n_2(t,\tau)\right]\cdot\text{conj}\left[s_{Rk\_part1}(t,\tau)+n_1(t,\tau)\right] \quad (32)$$

In the presence of noise, the expected value of $x(t,\tau)$ at the maximum point $\tau=\tau_0$ is

$$\begin{aligned}E\{\text{SDFC}_{s+n}(\tau_0)\} &= E\left\{\left[s_{Rk\_part2}(t,\tau)+n_2(t,\tau)\right]\cdot\text{conj}\left[s_{Rk\_part1}(t,\tau)+n_1(t,\tau)\right]\right\} \\ &= E\left\{s_{Rk\_part2}(t,\tau)s^*_{Rk\_part1}(t,\tau)+n_2(t,\tau)n_1^*(t,\tau)\right\} \\ &= \frac{\sigma_{0k}^2 B\tau}{4}\exp\left[-j\frac{4\pi}{c}\Delta f_r R_k(t)\right]+\frac{V^2}{2}\end{aligned} \quad (33)$$

The second-order moment is defined as

$$\begin{aligned}&E\left\{\left|\text{SDFC}_{s+n}(\tau_0)\right|^2\right\} \\ &= E\left\{\begin{bmatrix}s_{Rk\_part2}(t,\tau)+n_2(t,\tau)\end{bmatrix}\begin{bmatrix}s^*_{Rk\_part1}(t,\tau)+n_1^*(t,\tau)\end{bmatrix} \\ \begin{bmatrix}s_{Rk\_part1}(t,\tau)+n_1(t,\tau)\end{bmatrix}\begin{bmatrix}s^*_{Rk\_part2}(t,\tau)+n_2^*(t,\tau)\end{bmatrix}\right\} \\ &= E\left\{\begin{matrix}s_{Rk\_part2}(t,\tau)s^*_{Rk\_part2}(t,\tau)s^*_{Rk\_part1}(t,\tau)s_{Rk\_part1}(t,\tau) \\ +s_{Rk\_part2}(t,\tau)s^*_{Rk\_part2}(t,\tau)n_1^*(t,\tau)n_1(t,\tau) \\ +s_{Rk\_part2}(t,\tau)s^*_{Rk\_part1}(t,\tau)n_2^*(t,\tau)n_1(t,\tau)+s_{Rk\_part1}(t,\tau)s^*_{Rk\_part2}(t,\tau)n_1^*(t,\tau)n_2(t,\tau) \\ +s_{Rk\_part1}(t,\tau)s^*_{Rk\_part1}(t,\tau)n_2^*(t,\tau)n_2(t,\tau)+n_2^*(t,\tau)n_2(t,\tau)n_1^*(t,\tau)n_1(t,\tau)\end{matrix}\right\}\end{aligned} \quad (34)$$

By using the properties of the moments of complex Gaussian random variables [35], we can obtain

$$E\left[ s_{Rk\_part2}(t,\tau)s^*_{Rk\_part2}(t,\tau)s^*_{Rk\_part1}(t,\tau)s_{Rk\_part1}(t,\tau) \right] = \left(\frac{\sigma_{0k}^2 B\tau}{4}\right)^2$$

$$E\left[ s_{Rk\_part2}(t,\tau)s^*_{Rk\_part2}(t,\tau)n_1^*(t,\tau)n_1(t,\tau) \right] = \frac{\sigma^2}{2} E\left[ s_{Rk\_part2}(t,\tau)s^*_{Rk\_part2}(t,\tau) \right] = \frac{\sigma_{0k}^2 B\tau V^2}{8}$$

$$E\left[ s_{Rk\_part2}(t,\tau)s^*_{Rk\_part1}(t,\tau)n_2^*(t,\tau)n_1(t,\tau) \right] = \frac{\sigma^2}{2} E\left[ s_{Rk\_part2}(t,\tau)s^*_{Rk\_part1}(t,\tau) \right]$$
$$= \frac{\sigma_{0k}^2 B\tau V^2}{8} \exp\left[ -j\frac{4\pi}{c}\Delta f_r R_k(t) \right] \quad (35)$$

$$E\left[ s_{Rk\_part1}(t,\tau)s^*_{Rk\_part2}(t,\tau)n_1^*(t,\tau)n_2(t,\tau) \right] = \frac{\sigma^2}{2} E\left[ s_{Rk\_part1}(t,\tau)s^*_{Rk\_part2}(t,\tau) \right]$$
$$= \frac{\sigma_{0k}^2 B\tau V^2}{8} \exp\left[ j\frac{4\pi}{c}\Delta f_r R_k(t) \right]$$

$$E\left[ s_{Rk\_part1}(t,\tau)s^*_{Rk\_part1}(t,\tau)n_2^*(t,\tau)n_2(t,\tau) \right] = \frac{\sigma^2}{2} E\left[ s_{Rk\_part1}(t,\tau)s^*_{Rk\_part1}(t,\tau) \right] = \frac{\sigma_{0k}^2 B\tau V^2}{8}$$

$$E\left[ n_1^*(t,\tau)n_2(t,\tau)n_2^*(t,\tau)n_2(t,\tau) \right] = \frac{V^4}{2}$$

Combining (33), (34) and (35), we obtain the variance

$$\text{var}\{\text{SDFC}_{s+n}(\tau_0)\} = \frac{\sigma_{0k}^2 B\tau V^2}{4} + \frac{V^4}{4} \quad (36)$$

By inserting this expression in (16), and considering that the maximum of the signal alone is $\sigma_{0k}^2 B\tau/4$, we can express the output SNR after sub-band dual-frequency conjugate processing as

$$\text{SNR}_{\text{SDFC}} = \frac{\left(\dfrac{\sigma_{0k}^2}{4} B\tau\right)^2}{\dfrac{V^4}{4} + \dfrac{\sigma_{0k}^2}{4} B\tau V^2} = \frac{\text{SNR}_{\text{PC}}^2}{4 + 4\text{SNR}_{\text{PC}}} \quad (37)$$

where the input SNR and the SNR after range compression are denoted as $\text{SNR}_{\text{in}} = \sigma_{0k}^2/V^2$ and $\text{SNR}_{\text{PC}} = B\tau \cdot \text{SNR}_{\text{in}}$, respectively.

## References

[1] G. Sun, M. Xing, X. Xia, Y. Wu, and Z. Bao, "Robust ground moving target imaging using Deramp-keystone processing," *IEEE Trans. Geosci. Remote Sens.*, vol. 51, no. 2, pp. 966–982, Oct. 2013.

[2] J. Qian, X. Lv, M. Xing, L. Li, and Z. Bao, "Motion parameter estimation of multiple ground fast-moving targets with a three-channel synthetic aperture radar," *IET Radar Sonar Navig.*, vol. 5, no. 5, pp. 582-592, Jun. 2011.

performance," *IEEE Trans. Signal Process.,* vol. 59, no. 8, pp. 3576–3591, Aug. 2011.

[20] S. Luo, G. Bi, X. Lv, and F. Hu, "Performance analysis on Lv distribution and its applications," *Digital Signal Process.,* vol. 23, no. 3, pp. 797–807, May 2013.

[21] S. Q. Zhu, G. S. Liao, D. Yang, and H. H. Tao, "A new method for radar high-speed maneuvering weak target detection and imaging," *IEEE Geosci. Remote Sens. Lett.*, vol. 11, no. 7, pp. 1175-1179, Jul. 2014.

[22] J. Tian, W. Cui, and S. Wu, "A novel method for parameter estimation of space moving target," *IEEE Geosci. Remote Sens. Lett.*, vol. 11, no. 2, pp. 389–393, Feb. 2014.

[23] R. P. Perry, R. C. Dipietro, and R. L. Fante, "SAR imaging of moving targets," *IEEE Trans. Aerosp. Electron. Syst.*, vol. 35, no. 1, pp. 188–200, Jan. 1999.

[24] H. Ruan, Y. H. Wu, X. Jia, and W. Ye, "Novel ISAR imaging algorithm for maneuvering targets based on a modified keystone transform," *IEEE Geosci. Remote Sens. Lett.*, vol. 11, no. 1, pp. 128-132, Jan. 2014.

[25] S. Q. Zhu, G. S Liao, H. H Tao, and Z. W. Yang, "Estimating ambiguity-free motion parameters of ground moving targets from dual-channel SAR sensors," *IEEE J. Sel. Topics Appl. Earth Observ. Remote Sens.*, vol. 7, no. 8, pp. 3328-3349, Aug. 2014.

[26] J. Tian, W. Cui, Q. Shen, Z. X. Wei and S. Wu, "High-speed maneuvering target detection approach based on joint RFT and keystone transform," *Sci. China Proc. F.*, vol. 56, no. 6, pp. 1-13, Jun. 2013.

[27] S. Barbarossa, "Analysis of multicomponent LFM signals by a combined Wigner- Hough transform," *IEEE Trans. Signal Process.*, vol. 43, no. 6, pp. 1511–1515, Jun. 1995.
25

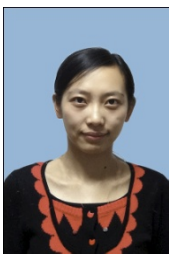

Jing Tian was born in Shandong, China, in November 1984. She received the B.S. and Ph.D. degrees in electrical engineering from Xidian University and Beijing Institute of Technology in 2006 and 2014, respectively. She is currently working as post-doctor in the second academy of China aerospace, Beijing.

Her research interests include moving-target detection, parameter estimation and imaging.

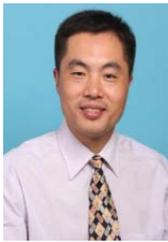

Wei Cui was born in Inner Mongolia Municipality, China in 1976. He received the Ph.D. degrees in electronics engineering from Beijing Institute of Technology in 2003. From 2003 to 2005, he worked as post-doctor in Radar Research Institute in Beijing Institute of Technology, where his research mainly concentrated on radar system and VLSI implementation of radar signal processing. Now, he worked as a professor and supervisor for doctorate students in Beijing Institute of Technology.

His research interests include space target detection and localization, array signal processing, and VLSI design.

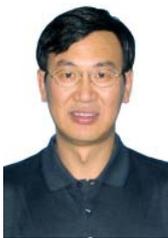

Siliang Wu was born in Anhui Province, China, in 1964. He received his Ph.D. degree from Harbin Institute of Technology in 1995 and then worked as a post-doctor in Radar Research Institute in Beijing Institute of Technology from 1996 to 1998. He is now a professor and supervisor for doctorate students in Beijing Institute of Technology and is a senior member of Chinese Institute of Electronics. His research interests include radar system and theory, satellite navigation and application of modern signal processing.